\documentclass[manuscript]{aastex701}
\usepackage{amsmath}

\begin{document}

\title{On the Divergent Evolution of Io and Europa as Primordial Ocean Worlds}

\author[orcid=0000-0000-0000-0001]{Yannis Bennacer}
\affiliation{Aix- Marseille Universit\'e, CNRS, CNES, Institut Origines, LAM, Marseille, France} \email{yannis.bennacer@lam.fr}

\author[orcid=0000-0000-0000-0002]{Olivier Mousis} 
\affiliation{Solar System Science and Exploration Division, Southwest Research Institute, 1301 Walnut St, Ste 400, Boulder, CO, USA}
\email{olivier.mousis@swri.org}

\author[orcid=0000-0000-0000-0001]{Vincent Hue}
\affiliation{Aix- Marseille Universit\'e, CNRS, CNES, Institut Origines, LAM, Marseille, France} 
\email{vincent.hue@lam.fr}

\begin{abstract}

The Galilean moons exhibit a decrease in bulk density with distance from Jupiter, which may reflect differences in evolutionary paths and water loss. Early in its history, Jupiter was more luminous which may have driven substantial atmospheric escape on Io and Europa. We investigate whether Io could have lost its water inventory while Europa retained its volatiles, assuming both moons initially accreted hydrous silicates. The formation and early thermal evolution of the protosatellites are modeled using an interior evolution model coupled with an atmospheric escape framework. Dehydration timescales and volatile losses for Io and Europa are computed during their early evolution, accounting for accretional heating from both satellitesimal and pebble accretion and irradiation from Jupiter’s primordial luminosity. Europa likely retained most of its volatiles under nearly all plausible formation and evolution scenarios, as large-scale dehydration would have taken place only after the first $\sim$10 Myr of its evolution. In contrast, Io was unlikely to lose a substantial amount of water through atmospheric escape and therefore probably accreted predominantly anhydrous silicates. If Europa initially accreted hydrous minerals, the present-day volatile contrast between Io and Europa could be explained by their relative locations with respect to the phyllosilicate dehydration line in the Jovian subnebula. Distinct evolutionary pathways or atmospheric escape processes alone appear insufficient to reproduce the observed differences.

\end{abstract}

   \keywords{Galilean satellites - Satellite Formation - Atmospheric dynamics - Jupiter - Europa - Io
               }


\section{Introduction}
The striking variation in the ice-to-rock ratio among the Galilean moons is one of the Jovian system's most intriguing puzzles. Two main hypotheses have been put forward to explain this diversity. The first suggests that the moons'contrasting water contents reflect the thermal structure of Jupiter's circumplanetary disk (CPD) at the time of their formation. According to this theory, Io and Europa formed within the snowline, where high disk temperatures limited ice condensation, while Ganymede and Callisto formed further out, in regions cold enough for water ice to accumulate \citep{Lunine&Stevenson_1982, Canup&Ward_2002,Mosqueria&Estrada_2003_a,Ronnet_etal_2017}. An alternative scenario suggests that all four satellites initially formed as ocean worlds rich in water ice, and that the observed compositional gradient arose later through the loss of volatiles in the inner moons \citep{Bierson&Nimmo_2020}. In this framework, hydrodynamic atmospheric escape, driven by intense heating and low gravity, emerges as the most plausible mechanism for stripping water from Io and Europa. Such escape could have occurred during accretion, when disk heating and impact energy created transient surface oceans whose atmospheres were rapidly removed. Even after the CPD dissipated, Jupiter’s residual luminosity may have sustained high surface temperatures, potentially triggering a runaway greenhouse effect capable of depleting the inner moons’ primordial water inventories \citep{Bierson_etal_2023}.

If Europa did not retain its initial water, its present-day hydrosphere may have originated from the late release of water during the dehydration of hydrous silicates. This process is consistent with several independent studies \citep{Kargel_etal_2000, Melwani_Daswani_etal_2021, Trinh_etal_2023, Petricca_etal_2025}. By that time, Jupiter’s luminosity would have decreased, making the moon’s surface too cold to sustain a stable ocean-atmosphere system. This would have protected its reformed hydrosphere from further loss. However, for Io, the complete removal of any primordial ocean likely required rapid dehydration very early on, before Jupiter’s luminosity declined. No later process appears capable of efficiently removing an established hydrosphere; tidal heating alone would have been insufficient to eliminate an ice-rich shell \citep{Bierson&Steinbrugge_2020}.

Here, we investigate the early thermal evolution of Io and Europa to quantify their dehydration timescales and volatile mass loss. We explore the implications of these processes for Europa’s current water inventory and the potential presence of a primordial ocean on Io. Ganymede and Callisto are not included in this analysis because their higher surface gravities, cooler formation environments, and weaker tidal dissipation likely prevented significant atmospheric escape and volatile depletion.

\section{Methods}

The formation and early thermal evolution of the protosatellites are modeled using the interior evolution model of \citet{Bennacer_etal_2025} coupled with the hydrodynamic escape model of \citet{Lehmer_etal_2017}.

\subsection{Interior Evolution}

The thermal history is computed using a one-dimensional spherical heat diffusion equation with a time-dependent radius. Radiogenic heating is assumed to mainly arise from  $^{26}$Al, $^{60}$Fe, and $^{53}$Mn. Newly accreted material layers produce heat associated with their formation time, \(t + t_{\text{start}}\), where \(t_{\text{start}}\) marks the onset of accretion after the formation of calcium-aluminum inclusions (CAIs) in the PSN. The radiogenic heating rate, $Q(t)_{\text{rad}}$, is calculated following the approaches of \citet{Bennacer_etal_2025} and \citet{Trinh_etal_2023}, using the isotope concentrations, heat production rates, and decay constants listed in Table \ref{table:1}.



\begin{table}[h!]
\caption{Heating of the main radiogenically-active compounds}                 
\centering                        
\begin{tabular}{c c c c}      
\hline\hline               
 Isotope & $^{26}$Al & $^{60}$Fe & $^{53}$Mn \\        
\hline 
Concentration (ppb)         & 600           & 100                       & 25.7 \\                      
Heat Production (W/kg)      & 0.355         & $7.0 \times 10^{-2}$      & $2.7 \times 10^{-2}$ \\    
Decay constant (Myr)        & 0.716         & 1.50                      & 3.70    \\
\hline                                  
\end{tabular}
\label{table:1}    
\end{table}

The rate of tidal energy dissipation in tidally locked satellites is given by
\begin{equation}
\frac{dE}{dt} = \frac{21}{2} \frac{k_2}{Q} \frac{\omega^5 R_s^5 e^2}{G},
    \label{equation:tidal}
\end{equation}
\noindent where $k_2$ denotes the Love number, $Q$ the tidal quality factor, $e$ the orbital eccentricity, $G$ the gravitational constant, and $\omega$ the mean motion. Constant orbital parameters $(\omega, e)$ are assumed during the early evolutionary stages, in accordance with the formation of Io and Europa as part of the final generation of satellites exhibiting limited inward migration \citep{Canup&Ward_2006}. The current eccentricity of Io is applied to both satellites to prevent the overestimation of Io’s internal dissipation. This has a negligible influence on Europa.


Tidal deformability, expressed as $k_2/Q$, depends on a satellite’s internal structure and is poorly constrained. 
Assuming Io and Europa accreted from similar materials, we adopt Io’s present-day value, $k_2/Q \approx 1.5 \times 10^{-2}$ \citep{Lainey_etal_2009}, is adopted for both satellites. This choice reasonably reproduces the current levels of tidal dissipation in Io and Europa when formation is assumed to have occurred near their present orbital locations. It also allows us to explore a broader range of formation distances by varying the orbital frequency, $\omega = \sqrt{GM_{\text{Jup}}/a^3}$, where $a$ is the semi-major axis of the moon (see section \ref{subsection:parameters_space_exploration}).


At high temperatures, hydrous minerals dehydrate, releasing water and reducing the rock mass by an equivalent amount. We assume that antigorite begins to dehydrate at 873 K, consistent with experimental results \citep{Sawai_etal_2013} and with the findings of \cite{Wakita&Sekiya_2011}. Dehydration consumes latent heat, $L = 417$ kJ kg$^{-1}$ \citep{Holland&Powell1998, Wakita&Sekiya_2011}, leading to an energy loss per unit volume given by

\begin{equation}
Q_h = \rho L \frac{dX}{dt},
\end{equation}




\noindent where $\rho$ is the bulk density. The evolution of the hydrated rock mass, $m_h = X m$, is tracked, where $X$ denotes the hydrated mass fraction of rocks within a given layer. At the start of the simulations, all rocks are assumed to be fully hydrated ($X = 1$). The temporal evolution of $X$ is then expressed as 

\begin{equation}
X = \frac{X m - d m_h}{m - wd m_h},
\end{equation}

\noindent where $w$ represents the initial water mass fraction (wt\%) in hydrated minerals and $dm_h$ denotes the mass of rock undergoing dehydration during a time step. Hydrated silicates are assumed to contain 6.8 wt\% H$_2$O, consistent with the composition of investigated mineral assemblages \citep{McKinnon_etal_2017} and studies of icy bodies \citep{Trinh_etal_2023}. The released water is assumed to migrate rapidly toward the surface, contributing to the hydrosphere through buoyancy-driven transport \citep{Melwani_Daswani_etal_2021, Trinh_etal_2023}. 

\subsection{Accretion and Atmospheric Escape}

Satellites are assumed to grow by accreting hydrous minerals from the circumplanetary disk (CPD) Given the poorly constrained size distribution and impact frequency of incoming solids, we consider two end-member accretion mechanisms, satellitesimal accretion and pebble accretion.

In the first scenario (oligarchic growth), proto-satellites accrete kilometer-scale impactors that penetrate deeply and deposit energy within the interior. These large impacts are modeled stochastically using a Monte Carlo approach \citep{Bennacer_etal_2025}. A large impact generates a shock wave inward, concentrating peak pressures within an approximately spherical isobaric core. A fraction of the projectile’s kinetic energy ($\gamma = 0.2$) is converted into shock heating, raising the temperature there according to:

\begin{equation}
\Delta T = \frac{4\pi}{9} \frac{\gamma\rho G R_s^2}{h_m c_p},
\end{equation}  

\noindent where $h_m\sim5.8$ denotes the ratio of the effectively heated volume to that core region. Outside this zone, both shock pressure and thermal perturbations decay steeply, scaling as $\Delta T \left(\frac{r_{\rm ic}}{\bar{r}}\right)^m$ with $m=3.4$ (see \cite{Bennacer_etal_2025} for details).

In the second scenario (pebble accretion), the surface of proto-satellites is continuously bombarded by small impactors throughout accretion. These impactors are sufficiently small to justify a continuous accretion approximation and are assumed to share the local CPD temperature. In addition to the gravitational energy released upon impact, the surface is also heated or cooled depending on whether the impactors are warmer or colder than the surface. 

The mass inflow rate in both scenarios is given by

\begin{equation}
\dot M = 7.15 \left( \frac{M_s}{M_{\text{tot}}} \right)^{2/3} \frac{M_{\text{tot}} - M_s}{\tau_{\text{acc}}},
\end{equation}

\noindent where $M_s$ is the satellite mass and $M_{\text{tot}}$ the total available mass. The scaling $\dot M \propto M^{2/3}$ captures the growth behavior in both oligarchic and two-dimensional pebble accretion regimes \citep{Kokubo&Ida_1998, Ida_etal_2016}.

The time-dependent surface temperature $T_{\text{surf}}$ is determined by a global energy balance, including accretion of pebbles, radiative heating from the CPD, Jupiter and the Sun, mass loss and radiative cooling \citep{Kuramoto&Matsui1994,Bierson_etal_2023}:

\begin{equation}
\begin{split}
& \underbrace{\frac{GM_s \dot{M}}{R_s} - c_p \dot{M}(T_{\text{surf}} - T_e)}_{\text{Accretion}} 
+ \underbrace{\pi R_s^2(F_s + F_{\text{jup}} + 4\sigma T_e^4)}_{\text{Irradiation}} \\
&\quad = \underbrace{\sigma T_{\text{surf}}^4 4\pi R_s^2}_{\text{Radiative cooling}} 
+ \underbrace{4\pi R_s^2 \rho_s u_s \left(\frac{GM_s}{R_s} + L_v\right)}_{\text{Mass loss}},
\end{split}
\label{equation:Tsurf}
\end{equation}

\noindent where $\sigma$ is the Stefan-Boltzmann constant, $T_e$ the temperature of the surrounding disk and $R_s$ and $M_s$ are respectively the radius and the mass of the growing satellites. The left-hand side of the equation represents, from left to right, the gravitational potential energy delivered by pebbles, the heating (or cooling) induced by these particles, the absorbed stellar ($F_s$) and Jupiter ($F_{\text{jup}}$) flux, and radiation from the background disk $T_e$. We assumed an isothermal optically thin atmosphere ($\tau =0$) \citep{Lehmer_etal_2017} and a Bond albedo of zero because most of Jupiter's radiation peaks in the infrared. The atmospheric density at the surface level is $\rho_s$ and the velocity $u_s$. When volatiles start to escape, the energy flux lost is a contribution of the latent heat of vaporization ($L_v$), and the required energy flux to lift the molecules out of the gravity well ($\rho_s u_sGM_s/R_s$).

During Io’s and Europa’s early evolution, atmospheric escape was dominated by thermally driven hydrodynamic outflow. At surface temperatures above $\sim$200 K and under low gravity, our calculations yield Jeans parameters below $\lambda \approx 2$, marking the transition from molecular to collective, hydrodynamic flow \citep{Volkov_etal_2011}. In this regime, excess thermal energy at the base of the atmosphere drives a vertical hydrogen wind that accelerates to the sound speed at a critical point, with the corresponding atmospheric mass loss rate given by:

\begin{equation}
\dot M_{\text{loss}} = 4\pi R_s^2\rho_su_s.
\end{equation}

\noindent $\rho_s$ is obtained using the ideal gas law $p_s=\rho_s u_0^2$ where $p_s$ is the surface pressure and $u_0^2=kT/m$ the isothermal sound speed. As the surface ocean creates an atmosphere around the satellite in saturated vapor pressure (SVP) equilibrium, surface pressure is directly given by $p_s=p_{\text{sat}}(T)$, with $p_{\text{sat}}(T)$ the saturated pressure.

The isothermal planetary wind is derived assuming an isothermal atmosphere, which represents the maximum atmospheric loss rate and using the hydrodynamic equations of an ideal gas in the steady state \citep{Lehmer_etal_2017}: 
\begin{equation}
    (u^2-u_0^2)\frac{1}{u}\frac{du}{dr} = 2\frac{u_0^2}{r} -\frac{GM}{R_s^2}
\label{equation:isothermal_wind}
\end{equation}

From this equation, and by neglecting the subsonic solution $(du/dr=0)_c$, which would require a finite background pressure, the critical point can be directly inferred as $r_c = GM/2u_0^2$, where the vertical wind reaches the isothermal sound speed. Following the approach of \citet{Lehmer_etal_2017}, the term $u^2$ is neglected, and an analytic expression for the surface vertical velocity is obtained by integrating equation (\ref{equation:isothermal_wind}) from the surface $R_s$ to $r_c$:

\begin{equation}
u_s = u_0 \left(\frac{r_c}{R_s}\right)^2 \text{exp}\left[-\frac{1}{2} + \frac{GM}{u_0^2}\left(\frac{1}{r_c} - \frac{1}{R_s}\right)\right].
\end{equation}

Consequently, the mass loss rate depends solely on the surface gravity and surface temperature.

\section{Results}

\subsection{Model parameters}

To couple accretional heating with the disk’s thermal evolution, we adopt the one-dimensional gas-starved accretion disk model of \citet{Canup&Ward_2002}, in which the midplane temperature evolves exponentially as follows:

\begin{equation}
T_d(t) = T_{d,0}\text{exp}\left(\frac{-t}{4\tau_G}\right).
\label{equation:Tdisk}
\end{equation}

\noindent where $\tau_G \equiv M_J/\dot M_J$ is the inflow timescale onto Jupiter and $T_{d,0}$ is the initial midplane temperature that strongly depends on the CPD opacity $\kappa$, the timescale $\tau_G$, and the viscosity parameter $\alpha$ \citep{Canup&Ward_2002}. Term 4 in Eq.~\ref{equation:Tdisk} results from balancing viscous heating with blackbody cooling, yielding $T_d^4(t) \propto \exp(-t/\tau_G)$. For our CPD temperature profile, we use the same parameter values as in their Figure 5, i.e., $\kappa = 10^{-4}$ cm$^2$ g$^{-1}$, $\tau_G = 5\times 10^6$ yr, and $\alpha = 5\times 10^{-3}$.  

Our simulations begin at $t_{\text{start}}$, the onset of the accretion phase relative to CAI formation. Although $t_{\text{start}}$ is poorly constrained, lower limits have been proposed based on Callisto’s cold and undifferentiated state ($t_{\text{start}} \gtrsim 3$–3.5 Myr) \citep{Barr&Canup_2008,Bennacer_etal_2025}, as this parameter controls the abundance of short-lived radionuclides available during satellite formation. In this study, we set $t_{\text{start}} = 3$ Myr after CAI formation to explore the most favorable conditions for Io’s early dehydration. 

We assume that the CPD remains optically thick throughout the accretion phase and becomes optically thin once satellite growth ceases, a transition proposed to trigger rapid ice loss from Io and Europa \citep{Pollack&Reynolds_1974,Bierson_etal_2023}. This transition marks the onset of substantial irradiation of the satellites by Jupiter and is directly linked to the accretion duration, $\tau_{\text{acc}}$, treated as a free parameter ranging from several hundred thousand to a few million years. Jupiter’s luminosity is prescribed following \citet{Fortney_etal_2011}, a model valid roughly 1 Myr after Jupiter’s formation \citep{Marley_etal_2007}. In some simulations, our model allows satellite irradiation before this epoch, which may overestimate Jupiter’s luminosity. This, however, does not affect the results, as neither dehydration nor atmospheric mass loss occurs prior to 1 Myr (Figure~\ref{figure:example_model}).

\subsection{Case Study}

\begin{figure*}[t]
\centering
\includegraphics[width=0.95\textwidth]{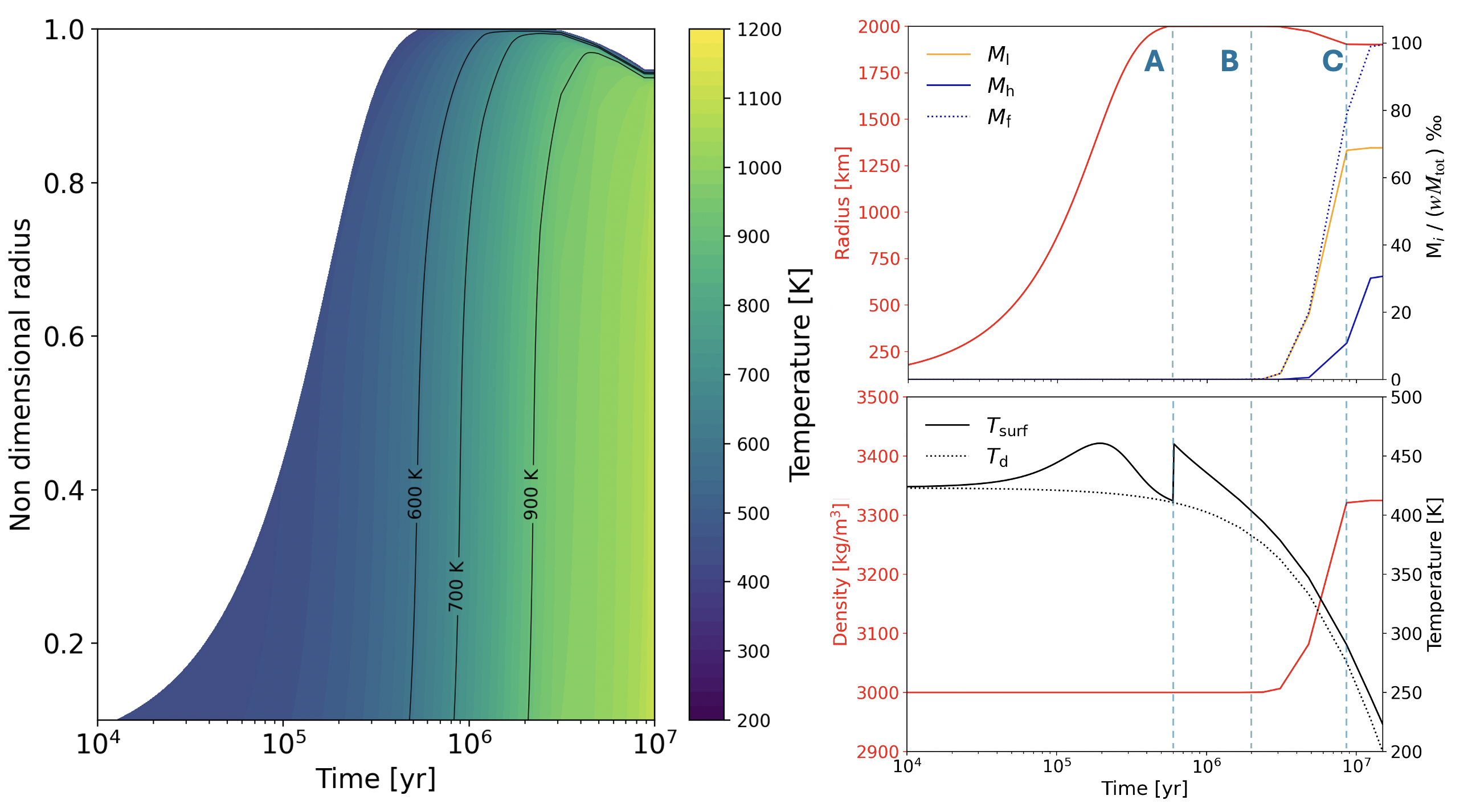}
\caption{Simulation of an Io--like protosatellite forming at $a = 6~R_{\text{Jup}}$. The body accretes only pebbles, growing from an initial radius $R_0 = 100$ km (at $t_{\text{start}} = 3$ Myr) to a final radius $R_f = 2000$ km over $\tau_{\text{acc}} = 0.6$ Myr. {\it Left}: evolution of the internal temperature profile. {\it Top right}: evolution of radius (red), mass loss $M_{\text{l}}$ (yellow), hydrosphere mass $M_{\text{h}}$ (blue), and fluid masses released from hydrous minerals $M_{\text{f}}$ (blue, dotted), all normalized to the total water mass fraction $w M_{\text{tot}}$ with $M_{\text{tot}}$ the total mass accreted. {\it Bottom right}: evolution of density (red), surface temperature (black), and disk temperature (dotted). Blue dashed lines and labels A, B, and C mark the successive model phases: disk dissipation, dehydration of hydrous minerals, and the end of atmospheric mass loss.}
\label{figure:example_model}
\end{figure*}

Figure \ref{figure:example_model} shows an example run of our model, where we compute the thermal evolution of protosatellites, the dehydration of minerals and the atmospheric mass loss. In this simulation, the protosatellite accretes only pebbles of hydrous minerals ($\rho$ $\sim$3000 kg/m$^3$) at Io’s orbital distance ($\sim$6 Jupiter radii) until reaching a final radius $R_f$ $\sim$2000 km. After formation ($\sim$ 6 $\times$ 10$^5$ yr.), Jupiter begins to irradiate the satellite (A), sharply increasing its surface temperature. The newly formed satellite continues to warm because of internal heating from radiogenic decay and tidal dissipation. At a temperature around 800 K ($\sim$2 Myr), hydrous minerals in the interior begin to dehydrate (B), releasing volatiles that progressively join the hydrosphere. With surface temperatures above the melting point of water, a water vapor atmosphere forms in equilibrium with a surface ocean, driving rapid material loss until the surface cools below the melting point ($\sim 9$ Myr). After this, the remaining water accumulates on the surface, forming an ice shell (C). This simulation produces a satellite with a final radius comparable to that of Io but with a lower density ($\rho \sim$3380 kg/m$^3$) and an ice shell several tens of kilometers thick.

\subsection{Sensitivity to Parameters}
\label{subsection:parameters_space_exploration}

\begin{figure*}[t]
\centering
\includegraphics[width=0.95\textwidth]{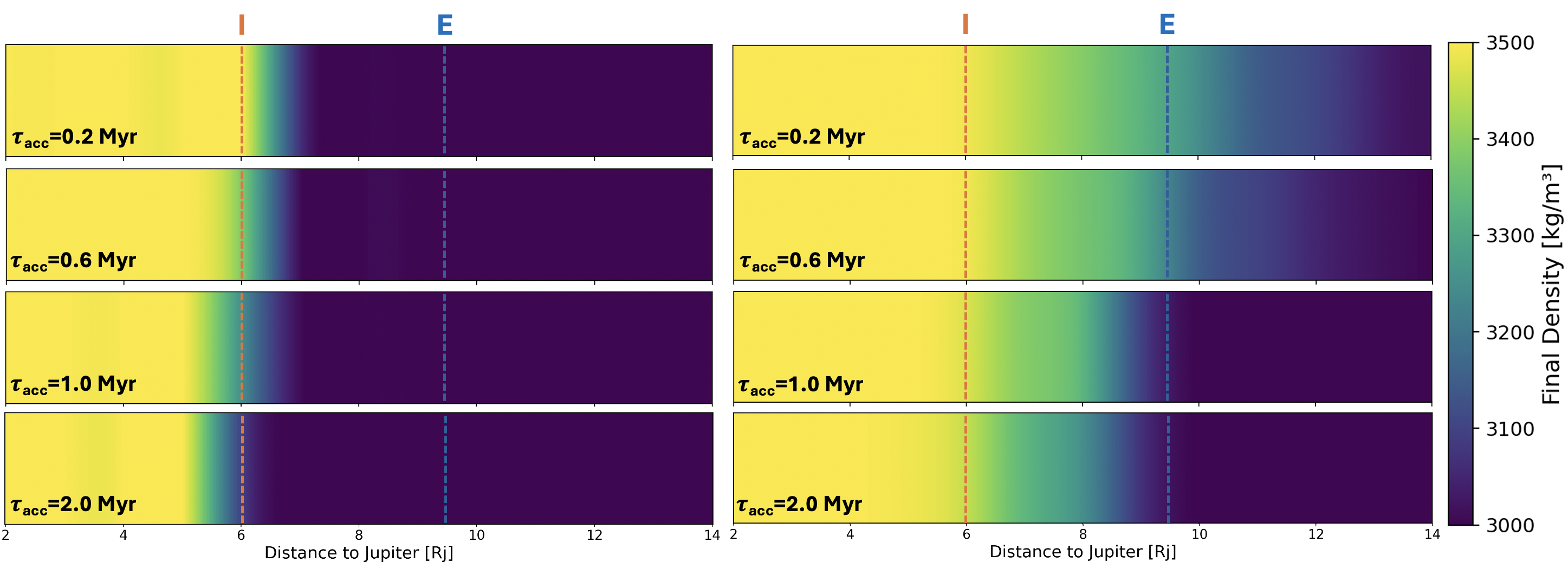}
\caption{Final density (color scale) of protosatellites at the end of the simulation for $R_f = 2000$ km, shown as a function of formation distance $a$ (sampled every 0.5 $R_{\text{Jup}}$ and interpolated using cubic interpolation) and accretion timescale $\tau_{\text{acc}}$. Accretion starts at $t_{\text{start}} = 3$ Myr, from either pebbles (left) or satellitesimals (right), each with a density of 3000 kg m$^{-3}$. Dotted lines indicate the present orbits of Io and Europa for reference.}
     \label{figure:parameter_exploration}
\end{figure*}

We test whether Io’s ($\sim$3500 kg m$^{-3}$) and Europa’s ($\sim$3000 kg m$^{-3}$) densities can be reproduced within our parameter space. Figure \ref{figure:parameter_exploration} shows the final protosatellite densities after atmospheric mass loss, as functions of formation distance $a$, accretion timescale $\tau_{\text{acc}}$, and impactor size distribution. A final radius of $R_f = 2000$ km is assumed for both satellites.

Volatile depletion results from the competition between dehydration ($\tau_{\text{deh}}$) and surface cooling ($\tau_{\text{cool}}$). Assuming pebble accretion, complete devolatilization occurs closer to Jupiter because of the  shorter dehydration timescales $\tau_{\text{deh}}$. The critical distance for full volatile loss  shifts inward from approximately 6 to 5 $R_{\text{Jup}}$ as $\tau_{\text{acc}}$ increases from 0.2 to 2 Myr, owing to reduced heating by short-lived radionuclides and decreased energy flux during slower accretion. On the other hand, the presence of large impactors, which deposit energy deeply within the interior, promotes earlier dehydration and weakens the dependence on $\tau_{\text{acc}}$. Consequently, a protosatellite accreting satellitesimals at or within Io’s current orbital distance would undergo complete volatile loss regardless of its accretion duration. As illustrated in Fig. \ref{figure:parameter_exploration}, Europa retains volatiles under most conditions, even if it formed near its present orbit. By contrast, reproducing Io’s fully devolatilized state ($\rho$ $\sim$3500 kg m$^{-3}$) requires at least one of the following conditions: (i) formation at $\lesssim 5R_{\text{Jup}}$, (ii) short accretion timescales ($\lesssim 0.4$ Myr), or (iii) accretion dominated by kilometer--scale impactors.

\section{Discussion}
\label{section:discussion}

The origin of the density gradient among the Galilean moons remains poorly constrained. This study explores the hypothesis that Io and Europa accreted hydrous silicates and identifies plausible ranges of model parameters that successfully reproduce the observed densities of the inner moons. One possible explanation is that the moons formed closer to Jupiter. However, this scenario is unlikely, as current models suggest they formed farther out, migrated inward through Type I migration \citep{Canup&Ward_2002}, and became locked in the Laplace resonance \citep{Peale&Lee_2002}, which prevented subsequent outward migration. A second possibility is that Io lost its water if it accreted predominantly pebbles over a short timescale ($\lesssim0.4$ Myr). Finally, Io and Europa may have accreted kilometer-sized satellitesimals, favouring Io's early dehydration and devolatilization (right panel Figure \ref{figure:parameter_exploration}).

Despite these plausible formation paths, our model shows that the window in which Io loses all its volatiles while Europa retains them is relatively narrow. Here we adopted several assumptions: (1) an optically thin disk, (2) an isothermal atmosphere, (3) an early onset of accretion ($t_{\text{start}} = 3.0$ Myr), and (4) we neglected the timescale needed for water resurfacing. Even under these favorable conditions, Io must have formed at its current orbit or closer to the gas giant. Io’s present density is difficult to reconcile with formation from hydrous silicates beyond its current orbit followed by Type I migration. The Galilean satellites may have avoided significant migration by forming late, after gas inflow waned \citep{Canup&Ward_2006}, but later formation would further delay dehydration and hinder volatile loss, making this scenario unlikely.

One might argue that a hotter CPD could yield a dry Io at larger orbital distances ($a$). This is unlikely, since internal heating and therefore the dehydration timescale depends strongly on tidal dissipation (Eq. \ref{equation:tidal}). A more plausible scenario is that elevated midplane temperatures dehydrated Io’s precursors before accretion, producing a dry moon \citep{Mousis_etal_2023}. 

The nominal formation onset in our model ($t_{\text{start}} = 3.0$ Myr), which sets the level of radiogenic heating, is constrained by Callisto’s partial differentiation \citep{Bennacer_etal_2025}. Slightly earlier formation is still viable in a pebble-accretion scenario ($t_{\text{start}} \gtrsim 2.5$ Myr) without driving Callisto to full differentiation \citep{Barr&Canup_2008, Shibaike_2025}. Additional simulations with $t_{\text{start}} = 2.5$ Myr produce similar outcomes (see Appendix Figure \ref{figure:pebble_2_5}): the moons’ properties can be reproduced, but Io cannot form beyond its present orbit. Furthermore, gas drag in the CPD would further reduce pebble velocities \citep{Shibaike_2025}, decreasing accretion heating and volatile loss.

Satellitesimal accretion, by contrast, cannot match Callisto’s differentiation for $t_{\text{start}} < 3$ Myr \citep{Bennacer_etal_2025}. Because this outcome depends on assumed hydrostatic pressure gradients, we also tested satellitesimal accretion with $t_{\text{start}} = 2.5$ Myr. In this case (Figure \ref{figure:satellitesimal_2_5}), Io and Europa can be matched with only modest migration ($\lesssim 2.5,R_{\text{Jup}}$), still far smaller than predicted in gas-starved or MMSN-type models \citep{Canup&Ward_2002, Mosqueira&Estra_2003_b}. However, such early formation would imply a thermally hot early evolution for Europa, inconsistent with its updated moment of inertia, which favors a colder history and a later formation time ($t_{\text{start}} \gtrsim 4.0$ Myr) \citep{Petricca_etal_2025}. This later timing is also consistent with Jupiter accreting material until 3–5 Myr after CAIs \citep{Kruijer_etal_2017, Kleine_etal_2020}.

To estimate the maximum possible mass loss from Io, we assumed an isothermal atmosphere, which represents the upper limit for escape. In practice, the escape rates for a water vapor atmosphere should fall between this isothermal limit and that of a saturated atmosphere \citep{Lehmer_etal_2017}. We also neglected greenhouse heating, since the model already assumes the highest possible mass-loss efficiency. As shown in Figure \ref{figure:example_model} (top right panel), volatiles that reach the surface are rapidly removed by hydrodynamic escape over much of the simulation, suggesting that further heating would have a negligible impact on the results.

We neglected atmospheric loss from sputtering by Jovian magnetospheric plasma and pickup ions, which today remove only $\sim10^{3}$ kg s$^{-1}$ from Io \citep{Wilson_etal_2002}. Removing even 1 wt\% of a primordial water inventory over 10 Myr would require sputtering rates $\gtrsim10^{5}$ kg s$^{-1}$. Since no studies indicate such elevated sputtering rates, we consider atmospheric sputtering negligible for removing an entire hydrosphere.

Despite the assumptions adopted in this work, Io was likely unable to lose its initial water inventory. After the dissipation of the accretion disk and the fading of Jupiter’s luminosity, the residual ice shell would not have been removed by tidal heating over geological timescales \citep{Bierson&Steinbrugge_2020}. This suggests that Io accreted primarily anhydrous silicates, and that the compositional contrast between the two inner moons reflects the thermodynamic structure of Jupiter’s CPD at the time of their formation, rather than divergent evolutionary or atmospheric loss processes. Io may have formed interior to the Phyllosilicate Dehydration Line (PDL) \citep{Mousis_etal_2023}, where water vapor released inside the PDL would diffuse outward and recondense near the snowline, producing the ice-rich Ganymede and Callisto.

An alternative scenario \citep{Bierson&Nimmo_2020} proposes that Io and Europa’s present-day characteristics resulted from atmospheric loss, assuming the satellites accreted a mixture of ice and rock. However, their CPD temperature profiles place Io well inside the snow line, where ice in small particles/pebbles would be unstable and likely sublimate before accretion. Inward delivery of water ice via larger satellitesimals could be envisioned, but melting water would likely induce rapid aqueous alteration \citep{Wakita&Sekiya_2011}, forming serpentine minerals before reaching the surface \citep{McKinnon&Zolensky_2003}. Consequently, volatile loss would occur only after dehydration. While a dedicated study considering both silicate alteration kinetics and water resurfacing is required before robust conclusions can be drawn, we expect that Io did not accrete significant ice or hydrous minerals, forming instead from dry, anhydrous material in the inner CPD.

In situ plume measurements and infrared spectroscopic observations from JUICE (PEP, SWI, MAJIS, UVS) and Europa Clipper (MASPEX) could provide key insights into the formation scenarios of the icy moons through determinations of their deuterium-to-hydrogen (D/H) ratios. In the scenario explored in this study, Europa is unlikely to have lost a significant fraction of its water inventory through atmospheric escape. Its D/H ratio is therefore expected to be comparable to that measured in hydrated asteroids and carbonaceous chondrites, and thus close to the terrestrial ocean value \citep{Alexander_etal_2012}.
For Ganymede and Callisto, the D/H ratios may be lower than that of Europa if the accreted water ice underwent isotopic exchange within the CPD during the outward diffusion of the vapor phase \citep{Horner_etal_2008,Mousis_etal_2023}, with the protosolar value representing the lower limit. However, as noted by \cite{Bierson&Nimmo_2020}, Europa could also acquire a higher D/H ratio than the outer satellites if it experienced substantial atmospheric mass loss. Future modeling efforts should quantify how water vapor diffusion within the CPD, together with isotopic fractionation, influences the resulting D/H ratio.

\begin{acknowledgements}

We thank the anonymous reviewer for their helpful comments. This research holds as part of the project FACOM (ANR-22-CE49-0005-01\_ACT) and has benefited from a funding provided by l'Agence Nationale de la Recherche (ANR) under the Generic Call for Proposals 2022. V. Hue acknowledges support from the French government under the France 2030 investment plan, as part of the Initiative d’Excellence d’Aix-Marseille Université – A*MIDEX AMX-22-CPJ-04. V.H. acknowledges the support of CNES to the Juno and JUICE missions. 

\end{acknowledgements}

%

\appendix

\setcounter{figure}{0}
\renewcommand{\thefigure}{A\arabic{figure}}

\begin{figure*}[t]
\centering
\resizebox{\hsize}{!}{\includegraphics[angle=0,width=4cm]{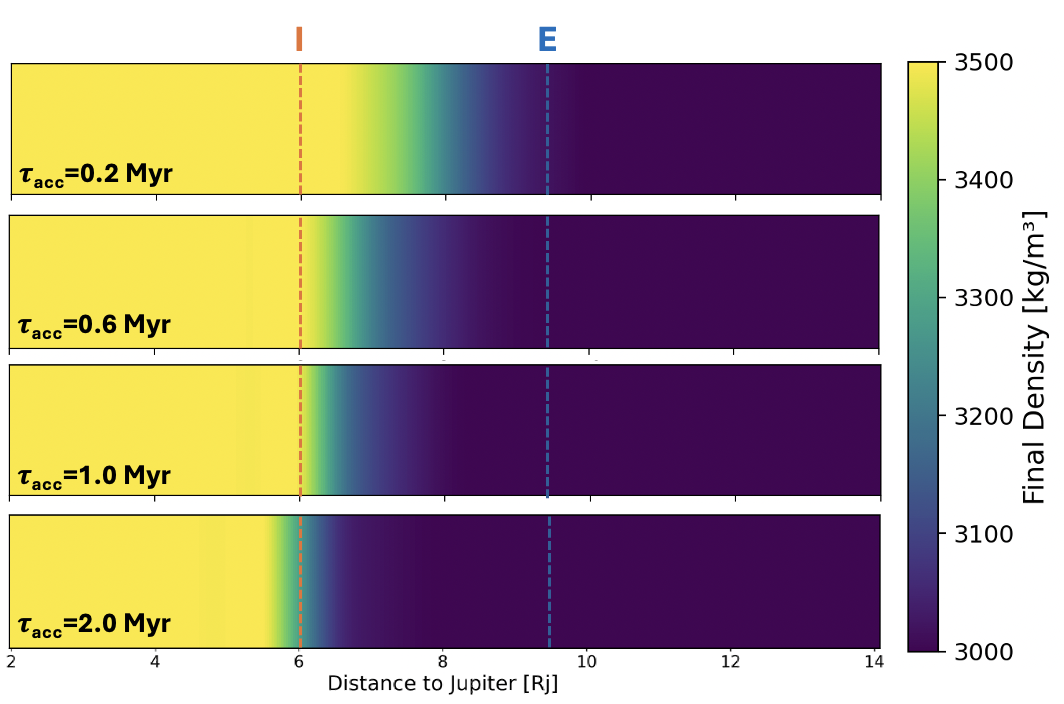}}
\caption{Final density of protosatellites accreting pebbles as a function of formation distance $a$ and accretion timescale $\tau_{\text{acc}}$ with $t_{\text{start}} = 2.5$ Myr.}
\label{figure:pebble_2_5}
\end{figure*}

\begin{figure*}[t]
\centering
\resizebox{\hsize}{!}{\includegraphics[angle=0,width=4cm]{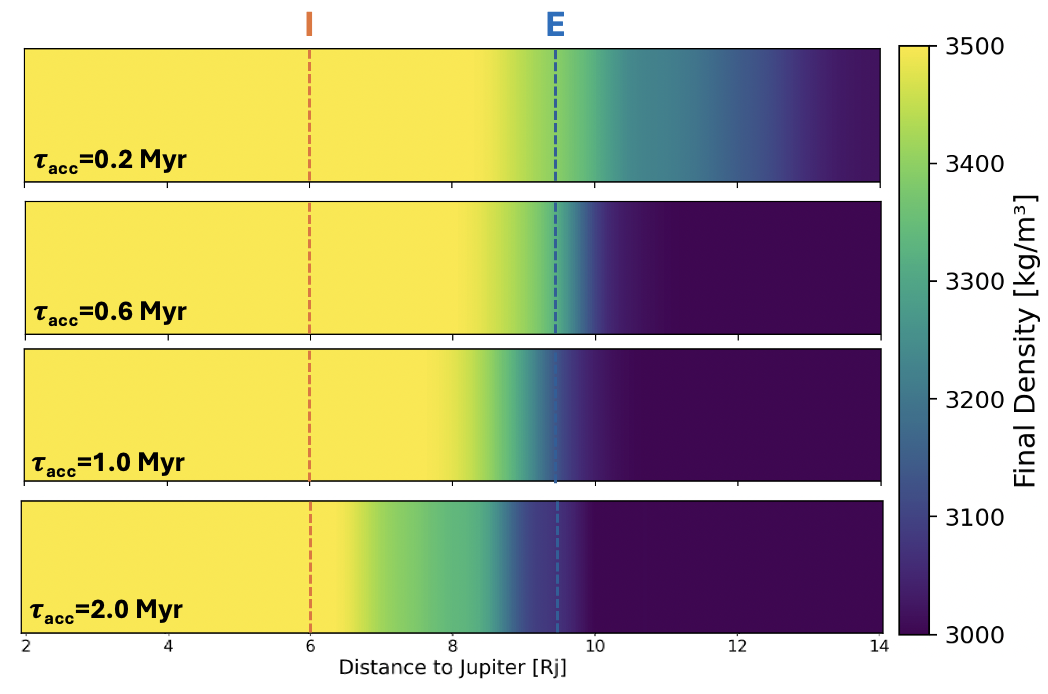}}
  \caption{Final density of protosatellites accreting satellitesimals as a function of formation distance $a$ and accretion timescale $\tau_{\text{acc}}$ with $t_{\text{start}} = 2.5$ Myr.}
     \label{figure:satellitesimal_2_5}
\end{figure*}

This Section presents additional simulations spanning a wider parameter range. Figures \ref{figure:pebble_2_5} and \ref{figure:satellitesimal_2_5} show the final protosatellite densities for an earlier formation onset, $t_{\text{start}} = 2.5$ Myr, for pebble and satellitesimal accretion, respectively.

\bibliography{sample701}{}
\bibliographystyle{aasjournalv7}



\end{document}